# Transition-metal phthalocyanine monolayers as new Chern insulators

Jie Li, Lei Gu and Ruqian Wu*

*Department of Physics and Astronomy, University of California, Irvine, California 92697-4575, USA.*

To explore new materials for the realization of the quantum anomalous Hall effect (QAHE), we studied electronic, magnetic and topological properties of transition-metal phthalocyanine (TMPc) monolayers in a square lattice. Many of them have large topologically nontrivial band gaps and integer Chern numbers. In particular, the Fermi level of MoPc lies right in the band gap ($E_g$=8.1 meV) and has large magnetic anisotropy energy (MAE=2.0 meV per Mo atom) and a Curie temperature of 16 K, showing its usefulness for applications. The presence of topologically protected edge state in a MoPc nanoribbon further confirms it as a new two-dimensional topological insulator. This much widens the search of QAHE materials for the design of quantum devices.

* E-mail: wur@uci.edu.

# 1. Introduction

Major technological innovations rely on the development of high performance and energy efficient microelectronic devices which often need novel functional materials. Topological insulators (TIs) and two-dimensional (2D) van der Waals (vdW) monolayers are recognized as the most promising materials for generating exotic properties such as the quantum spin Hall effect and the quantum anomalous Hall effect (QAHE). Bulk TIs behave not much differently from conventional insulators but their surfaces or edges have robust conducting states that are protected by the symmetry and have unique features such as spin-momentum interlock.[1-3] These stem from strong spin-orbit coupling (SOC) and peculiar electronic properties of materials such as band-inversion or the existence of Dirac states as for honeycomb lattices.[4] Many functional TIs have been proposed and fabricated, including $Hg_{1-y}Mn_yTe$ quantum wells,[5] graphene-based systems,[6-8] and organic molecular lattice.[9] In particular, magnetic TIs with a spin polarized gap may host spin polarized currents along their edges without dissipation, manifesting as the quantum anomalous Hall effect.[10] Despite extensive interdisciplinary endeavors, experimental realization of the QAHE appears to be still a challenge, as very few experimental observations of the QAHE have been reported to date.[2, 11-14] This is mainly hindered by the difficulties to control the distribution of magnetic atoms such as Cr and V via the doping approach, or to find appropriate magnetic insulators to magnetize yet not to damage the topological surface states of three dimensional TIs via the interfacial proximity.

Clearly, it is desired to search for new materials or different schemes for the realization of the robust QAHE and, furthermore, for the utilization in spintronic and quantum devices. Recently, a new series of 2D transition-metal phthalocyanine-based metal-organic material (TMPc) was synthesized.[15-17] TMPc materials have a planar 2D square lattice as shown in Fig. 1(a) and is transferable to other substrates,[18] which makes them attractive for studies of their magnetic and transport properties. As they are open in both sides, their effective SOC and magnetization can be conveniently tuned by filling different transition metal atoms in the cores, or by decorating them with adatoms or ligands. Therefore, TMPc monolayers may serve as ideal platforms for the studies of novel topological physics in non-honeycomb 2D materials, especially for the integration of topology and magnetism in a single system.

In this Letter, we determine the topological properties of a variety of TMPc lattices with the first-principles calculations and model analyses. Strikingly, a large topologically nontrivial band gap (~ 8 meV) is found for the MoPc lattice. This gap is spin polarized and sits right at the Fermi level, with an integer Chern number as we integrate the Berry curvature in the 2D Brillouin zone (BZ). Furthermore, we show that a one-dimensional MoPc ribbon has robust edge states. Therefore, the MoPc square lattice is predicted to be an excellent 2D material for the realization of the QAHE and has a potential for establishing new quantum phases such as Majorana states with a superconducting substrate. This paves a way for exploring novel 2D vdW topological materials with magnetic molecules.

## 2. Computational method

All of the first-principles calculations were carried out using the Vienna ab-initio simulation package (VASP) at the level of the spin-polarized generalized-gradient approximation (GGA) with the functional developed by Perdew-Burke-Ernzerhof (PBE).[19] The interaction between valence electrons and ionic cores was considered within the framework of the projector augmented wave (PAW) method.[20-21] The Hubbard U of 2.0 eV was adopted to describe the electron correlation in the d-shells of TM cores. The energy cutoff for the plane wave basis expansion was set to 500 eV. To sample the two-dimensional Brillouin zone, we used a 11×11 k-grid mesh. All atoms were fully relaxed using the conjugated gradient method for the energy minimization until the force on each atom became smaller than 0.01 eV/Å. The 1D band of the ribbon was calculated with a TB model based on the maximally localized Wannier functions (MLWFs), as implemented in Wannier90 code[22] and Wannier tools code.[23]

## 3. Results and discussion

The dashed square in Fig. 1(a) shows the supercell in our calculations for the TMPc square lattice which keeps the main body of a single transition metal-phthalocyanine molecule, i.e, the TM core, the metal-organic frame and the planer structure. Each supercell has 4 Hydrogen, 20 Carbon, 8 Nitrogen and 1 TM atoms to hold $D_{4h}$ point group symmetry. The optimized lattice constant is 10.68~10.81Å, (as shown in Table S1 in the Supplementary Information) which agrees with the experimental value (11±0.5 Å).[15] To further test the stability of TMPc monolayer without substrate, we take MoPc as an example and calculated the corresponding phonon band dispersion and baked them at 300K for 10ps by using ab initio molecular dynamics (AIMD) simulations as shown in Fig. S1 in the Supplementary Information. The absence of significant imaginary frequency in the phonon band dispersion and the total energy fluctuate around the equilibrium values without any sudden change and no structure destruction in AIMD simulations may indicates that these systems are stable. The central metal atom TM adopts the $TM^{2+}$ state charge and most of these systems are magnetic, e.g., with spin magnetic moments of 3.0, 4.0 and 2.0 $\mu_B$ for MnPc, MoPc and RuPc, respectively. To appreciate their electronic properties, the band structure of MoPc without and with SOC are given in Fig. 1(b). Interestingly, there is a pair of degenerate states at the M point in the spin-down channel when SOC is excluded in DFT calculations. This feature is desired for the search of large gap topological insulators, as we discussed previously.[24] Indeed, a band gap of ~8.1meV is opened when SOC is invoked. Importantly, the Fermi level also naturally resides in the gap which is perfect for the utilization of the topological feature without the need of gating. These key features retain in a reasonably large range of Hubbard U (0-3eV) for the Mo d-shell and by using the HSE06 calculations as shown in Fig. S2 and Fig. S3 in the Supplementary Information. The further calculations shows that the nontrivial band gap can be further enhanced to 15meV (35meV) by increasing the strength of SOC to 2 (10) times. In particular, the MoPc lattice is semiconducting in the spin-up channel and the spin intermix is negligible even as the SOC is included as seen in the mid panels in Fig. 1(b). Its states around the Fermi level have a perfect spin polarization and this material is hence ideal for spin transport applications.

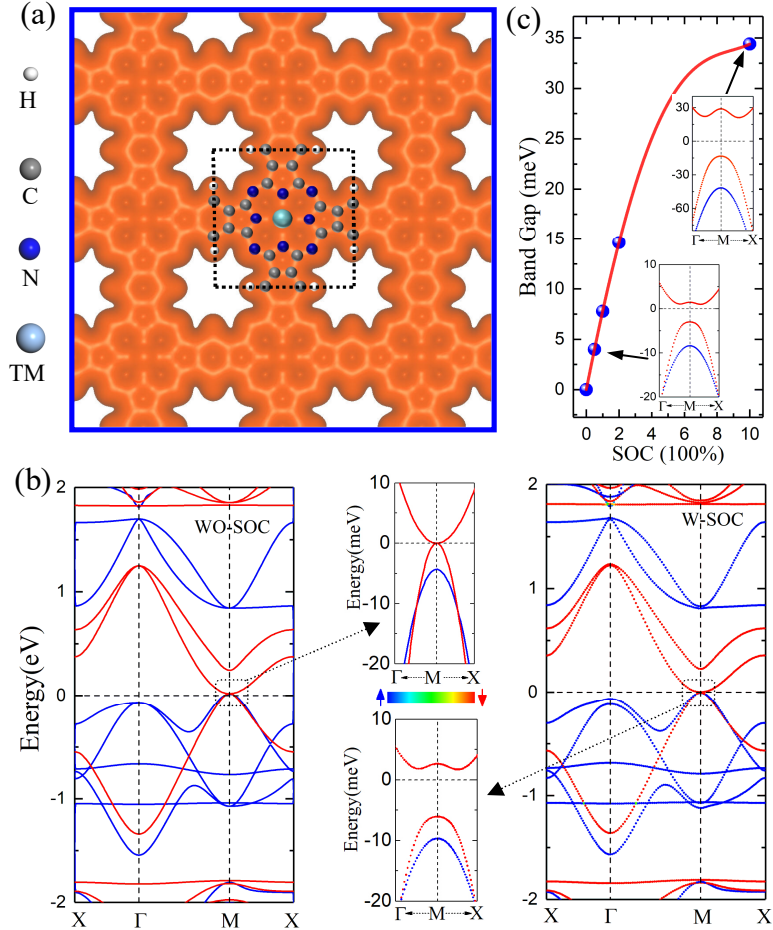

**Fig. 1.** (a) The schematic structure of 2D TMPc lattice (the dashed square shows the supercell for calculations), (b) The band structure of MoPc without (the left) and with (the right) SOC (the corresponding zoom-in band structures near the Fermi level around the M point are shown in the middle), (c) The band gap of MoPc as a function of the strength of SOC.

To check whether the SOC induced band gap is topologically nontrivial and gives rise to the QAHE, we calculated its Berry curvature $\Omega(k)$ in the whole Brillouin zone as shown in Fig. 2(a). The Berry curvature $\Omega(k)$ is defined by

$$\Omega(k) = 2Im \sum_{n\epsilon\{o\}} \sum_{m\epsilon\{u\}} \frac{\langle\psi_{nk}|v_x|\psi_{mk}\rangle\langle\psi_{mk}|v_y|\psi_{nk}\rangle}{(\varepsilon_{mk}-\varepsilon_{nk})^2},$$

Where $\{o\}$ and $\{u\}$ are the sets of occupied and unoccupied states; $\psi_{nk}$ and $\varepsilon_{nk}$ are the Bloch wave function and eigenvalue of the *n*th band at the k point; and $v_x$ and $v_y$ are the velocity operators, respectively. One may see that the Berry curvature around the M point has a large positive value and hence the gap is truly topologically nontrivial. The Chern number, C, which gives the Hall conductance as $\sigma_{xy} = C(e^2/\hbar)$, can be directly calculated by integrating the Berry curvature in BZ as

$$C = \frac{1}{2\pi}\int_{BZ} \Omega(k)d^2k$$

For discussions about the physical origin of the topological properties, we allow the Fermi level to vary in a range around its actual value, $E_F$, as shown in Fig. 2(b). It is obvious that a small terrace with C=1 presents as the Fermi level is placed in the SOC induced band gap. This is the hallmark of the QAHE, and thus the status of the MoPc lattice as a new 2D magnetic topological insulator is confirmed.

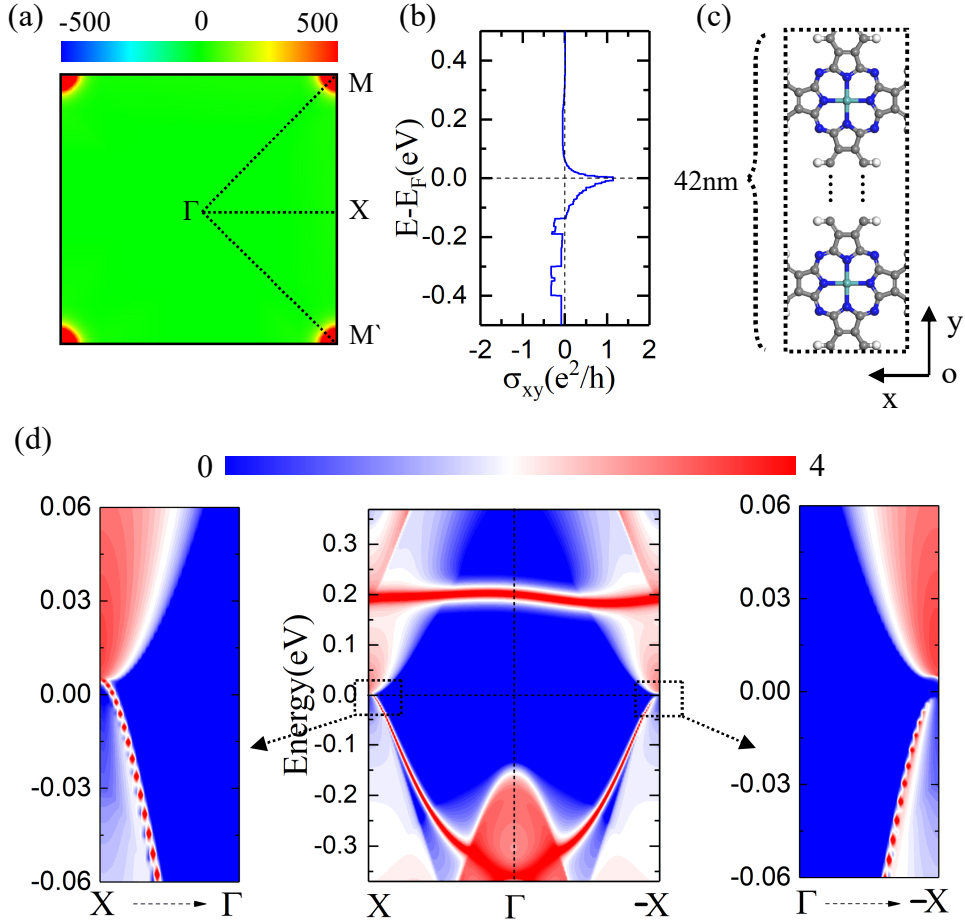

**Fig. 2.** (a) (b) Distribution of Berry curvature in the 2D Brillouin zone and Fermi level-dependent anomalous Hall conductance ($\sigma_{xy}$) for MoPc, (c) Geometry of MoPc nanoribbon periodic along the x-axis used in calculations. (d) The 1D band structure and the top edge states of this ribbon.

To further verify this, we construct a 1D MoPc nanoribbon about 42nm in width as shown in Fig. 2(c), which keeps the periodicity along the x-axis. The corresponding 1D band structure of this ribbon is calculated by using the maximally localized Wannier functions. As shown in Fig. 2(d), one top edge band is produced within the small band gap around the X point in the 1D BZ, to where the M-X-M line in the 2D BZ collapses. The top edge state connects the conduction band and valence band of the 2D lattice, as occurs in all topological materials. Meanwhile, the band gap keeps around the -X point, which is accord with the above Chern number (C=1). The opposite situation for the bottom edge state as shown in Fig. S4 in the Supplementary Information. Therefore, we confirm again that the SOC induced band gap for the 2D MoPc lattice is topologically nontrivial and has almost full spin-polarization.

Comparing to most topological insulators with the hexagonal honeycomb lattice, the square MoPc lattice is a non-Dirac electron system. Many non-Dirac electron systems have been found in germanene,[25] stanene,[13] and others semiconductor materials,[14,26-28] whose nontrivial topological effects are mostly associated with the p/d-like band inversion. The best scenario is to have p/d-like conduction band and valence band touching each other right at the Fermi level when the SOC effect is excluded. These bands become inversed after considering the intrinsic SOC, and a topologically nontrivial band gap opens. In order to understand the topological mechanism for the MoPc lattice, we reconstructed its band structure using the tight binding (TB) method. As shown in Fig. S5, the density of states of MoPc near the Fermi level consist of contributions from the $p_z$ orbitals of C and N atoms and $d_{xz/yz}$ orbitals of Mo. The same projections can be found by using the Wannier90 package as shown in Fig. S6 in which only the $p_z$ orbitals of C and N and $d_{xz/yz}$ orbitals of Mo are adopted as projected Wannier functions, and the fitted bands follow their DFT counterparts very well near the Fermi level. Thus, we may build up a mini tight binding model with $D_{4h}$ symmetry as shown in Fig. 3(b), where the $d_{xz}$ and $d_{yz}$ orbitals at the center and two types of $p_z$ orbitals on the edges and diagonals positions, respectively. The corresponding low-energy TB Hamiltonian contains two parts

$$H = H_0 + H_{soc}$$

with

$$H_0 = \sum_{i,\alpha} \epsilon_i^\alpha c_i^{\alpha+} c_i^\alpha + \sum_{<i,j>,\alpha,\beta} t_{i,j}^{\alpha,\beta} \left( c_i^{\alpha+} c_j^\beta \right) + h.c$$

and

$$H_{SOC} = -i\lambda \left( c_i^{dxz+} c_j^{dyz} - c_i^{dyz+} c_j^{dxz} \right) s_z + h.c$$

The former is spinless Hamiltonian parts and the latter is on-site SOC term. Where $\epsilon_i^\alpha$, $c_i^{\alpha+}$ and $c_i^\alpha$ represent the on-site energy, creation and annihilation operators for electron at the α-orbital of the *i*-th atom, respectively. $t_{i,j}^{\alpha,\beta}$ are the hopping parameters of electron between α-orbital of the *i*-th atom and β-orbital of the *j*-th atom. And λ is the atomic SOC strength of Mo and $s_z$ is the Pauli matrix (the detail as describe in the Supplementary Information). It turns out that the key features of the DFT bands near to the Fermi level are well reproduced as shown in the Fig. 3(a). Importantly, the degenerate bands disappear when we 1) change $d_{xz}$/$d_{yz}$ to other d orbitals such as $d_{xy}$; 2) change $p_z$ orbitals to $p_x$/$p_y$ orbitals; or 3) devastate its $D_{4h}$ symmetry by adding a splitting between $d_{xz}$ and $d_{yz}$ orbitals. It shows that the special orbital compositions ($p_z$, $d_{xz/yz}$ orbitals) and structural symmetry ($D_{4h}$) are essential for the topological band features around the M point. As the on-site SOC leads to the band gap opening, the mechanism is similar to the model proposed in our previous work.[24] As VBM and CBM of the MoPc lattice touch at Fermi level in the absence of SOC, the SOC is best used for producing a large-gap TI, as above DFT calculations (cf. Fig. 1c). The nontrivial band gap can be further enhanced by increasing the strength of SOC, which suggest a good strategy for tuning the topological band gap with different 4d or 5d TM atoms.

Results of other 3d, 4d and 5d TMPcs are shown in Fig S8 and some of them are interesting. For example, RePc lattice has a spin-polarized topological band gap as large as 72 meV. As the Fermi level does not lie in the band gap, additional treatments such as adsorbate decoration or high bias are needed to shift the Fermi level. In fact, most of these systems may manifest the QAHE if the position of the Fermi level can be shifted into the corresponding topological band gap. Obviously, this involves delicate technical issues and more research is needed to develop them for practical use.

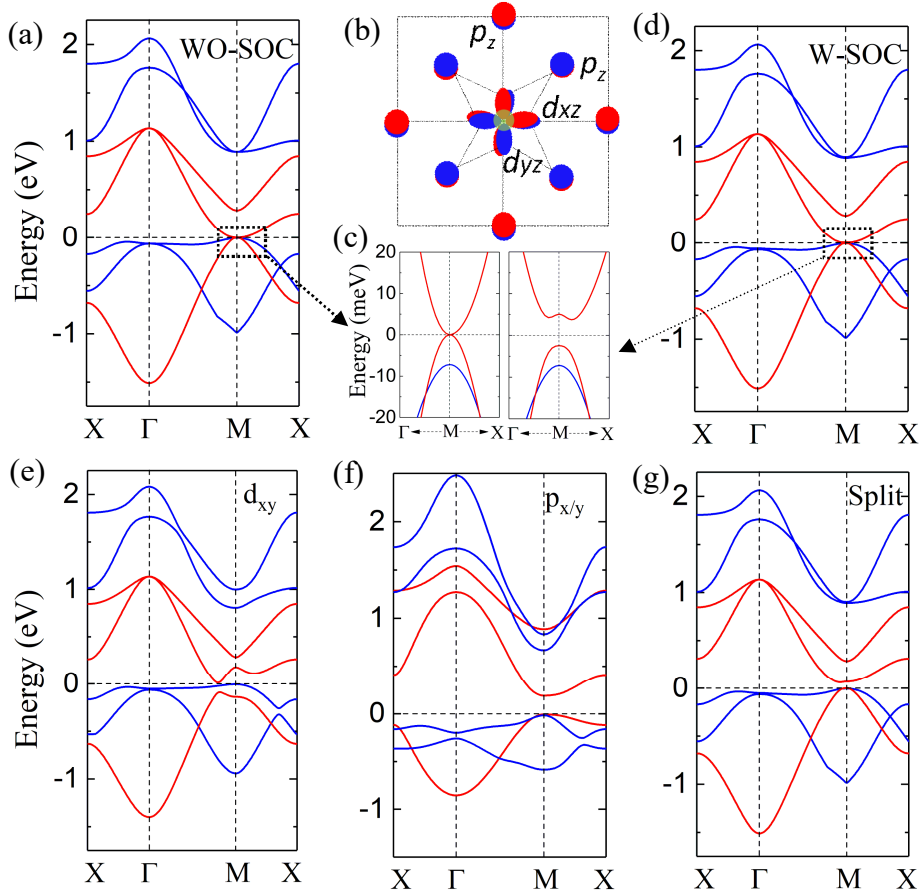

**Fig. 3.** (a) (d) The band structure of the TB model without and with SOC. Projected density of states of MoPc in near the Fermi level and spatial distributions of states at the tips of VBM and CBM. (c) The corresponding zoom-in band structures near the Fermi level around the M point. (b) Schematic structure of TB model with $d_{xz}$ and $d_{yz}$ orbitals at the central and six $p_z$ orbitals in the nearest and next nearest sites. (e)-(g) The case based on $p_z$ and $d_{xy}$ orbitals, the case based on $p_{x/y}$ and $d_{xz/yz}$ orbitals and the case based on $p_z$ and $d_{xz/yz}$ orbitals with splitting between $\varepsilon^{d_{xz}}$ and $\varepsilon^{d_{yz}}$ from minimal TB model, respectively.

To date, few experimental observations of QAHE were reported in extreme low temperature below 0.1K. There are many factors for this scenario, such as small topological insulator (TI) gap, magnetic disordering, and inhomogeneity of magnetization. With a naturally ordered lattice and a large TI gap of about 8.1 meV, the MoPc monolayer appears to avoid these factors. As the magnetic ordering sensitively depends on the magnetic anisotropy energy (MAE), one more piece of the puzzle is the direct determination of MAE of this system. Here, we use torque method and calculate the Fermi level dependent total and spin-decomposed MAEs as shown in Fig. 4(a). Large total MAEs ~ 2 meV (~ 20K) in a broad energy range around the actual Fermi level ($E_f^0$) mostly come from the cross-spin contributions, indicate a stable out-of-plane magnetization at a reasonably high temperature. Meanwhile, the corresponding Curie temperature ($T_C$) was calculated using the renormalized spin-wave theory (RSWT). [29-30] A reasonably high $T_C$ (16 K) was shown in the inset of Fig. 4(a), which offers a convenient condition for experimental observation. In addition, we also explore the strain effect on the topological band gap and the results are shown in the Fig. 4(b). One may see that the band gap is almost a linear function of the lateral strain and can be enlarged to more than 10 meV when the lattice is compressed.

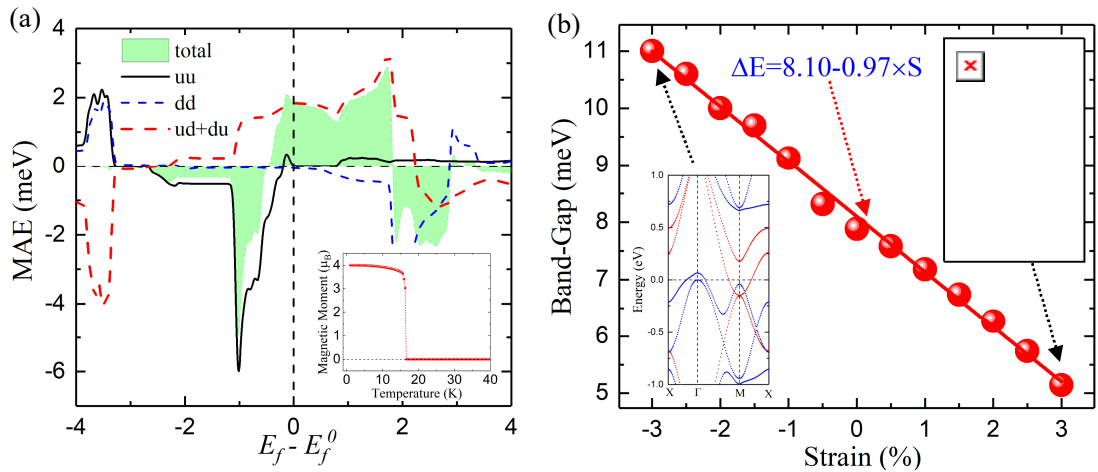

**Fig. 4.** (a) Fermi level dependent total and decomposed MAEs of MoPc from rigid band model (the inset is the magnetic moment as a function of temperature), (b) Band gap as a function of in-plane biaxial strain.

## 4. Conclusion

In summary, we investigated the electronic, magnetic and topological properties of TMPc-monolayers and found that the MoPc lattice is a new 2D magnetic topological insulator with a large nontrivial band gap of 8.1 meV. Direct calculations of the Chern number and edge state confirm it as a promising candidate for the realization of the QAHE. The topological band gap can be further expanded by applying a compressive lateral strain or by choosing heavier transition metal core. These results suggest a practical avenue for the realization of the QAHE in an easier lab condition and may have a significant impact on searching new topotronic materials.

## Conflicts of interest

The authors declare no competing financial interest.

## ACKNOWLEDGMENTS

Work was supported by DOE-BES (Grant No. DE-FG02-05ER46237). Calculations were performed on parallel computers at NERSC.